\documentclass[runningheads]{llncs}
\usepackage{graphicx}
\begin{document}
\title{One Size Fits All: A Conceptual Data Model for Any Approach to Terminology}
%
%
\author{Giorgio Maria Di Nunzio\inst{1,2} \and
Federica Vezzani\inst{3}}
%
%
\institute{Department of Information Engineering, University of Padova, Italy \and
Department of Mathematics, University of Padova, Italy \and
Department of Linguistic and Literary Studies, University of Padova, Italy\\
\email{\{giorgiomaria.dinunzio,federica.vezzani\}@unipd.it}}
\maketitle       
\begin{abstract}
In this paper, we want to speculate about the possibility to model all the currently known/proposed approaches to terminology into a single schema. We will use the Entity-Relationship (ER) diagram as our tool for the conceptual data model of the problem and to express the associations between the objects of the study. We will analyse the onomasiological and semasiological approaches, the ontoterminology paradigm, and the frame-based model, and we will draw the consequences in terms of the conceptual data model. The result of this discussion will be used as the basis of the next step of the data organization in terms of standardized terminological records and Linked Data.

\end{abstract}

\section{Introduction}

The use of standards is a fundamental step for reusability of digital resources. In Terminology, the reference ISO standard ISO 30042:2019,\footnote{ISO 30042:2019 Management of terminology resources - TermBase eXchange (TBX), \url{https://www.iso.org/standard/62510.html}} describes an XML-based terminology exchange format, designed to make terminology databases easier and safer to maintain, distribute, and use. Nevertheless, this standard is not immediately useful when a researcher want to expose the data structured as Linked Data.
Attempts to fill this gap has been proposed in the literature \cite{dibuono_et_al_2020a,dibuono_et_al_2020b}; however, there are still some manual corrections or interventions that are needed in order to make the conversion perfect~\cite{reineke_2014}, as well as problems in the depth of the information that can be represented with other models more concept-oriented~\cite{reineke_romary_2019}.

In this work, we want to discuss the problem from a different perspective. We make a step back and look at these issues as a data modeling problem. Our hypothesis is that if the terminological data are modeled correctly at an abstract level, then the data themselves can be exported or exposed in any format and maintain both the linguistic dimension and its relationship to the conceptual dimension. For this purpose, we will use the Entity-Relationship (ER) diagram as our tool for the conceptual model of the problem and the explanation of the associations between the object of the study. 

The paper is organized as follows: in Section~\ref{sec:conceptual} we introduce the basic elements of the conceptual data modeling and the pictorial representation that we use, the Entity-Relationship (ER) diagram. In Section~\ref{sec:model}, we analyze step by step four approaches in Terminology, onomasiological, semasiological, ontoterminological, frame-based, and we design the corresponding conceptual data model. In Section~\ref{sec:final}, we give our final comments and a view on the current and future work.

\section{Conceptual Data Modeling: ER Diagrams}\label{sec:conceptual}

\begin{figure}[t]
\begin{center}

\includegraphics[width=\textwidth]{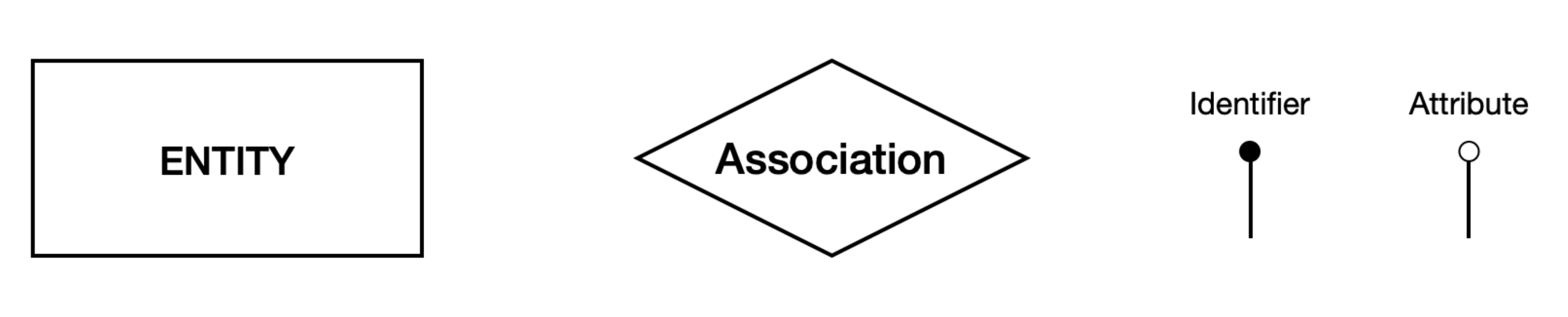}

\caption{ER diagram basic constructs: entity, association, attributes}
\label{fig:er_schema}
\end{center}
\end{figure}

A conceptual data model represents the objects of the domain of interest and the relationships among them. It is a (not-so) complicated visual language which is effective for communicating and describing how the knowledge is organized at an abstract level without any specific reference to how the data will be physically implemented. The Entity-Relationship (ER) Model~\cite{chen_1976} is a model of this kind, and its pictorial representation is called ER diagram. This diagram uses three basic elements, shown in Figure~\ref{fig:er_schema} that describes all the important semantic information about the domain of interest: entities, associations, attributes. An entity is the class of elements that are described with a set of attributes, some of the attributes are `identifiers' (distinguished with a black dots from the non-dientifiers) and distinguish uniquely each element of the class.  An association describes a class of elements that represent the relationships between entities.

To make an example, let us suppose to model the following problem described by Roche~\cite{roche_2012}:
\begin{quote}
[...] unlike classical terminology ontoterminology preserves the diversity of language between different communities of practice since they share the same domain and standardised conceptualisation. In point of fact, two different terms can denote the same concept whose identifier should be built so that we understand the right place of the concept in the ontology.
\end{quote}

\begin{figure}[t]
\begin{center}

\includegraphics[width=\textwidth]{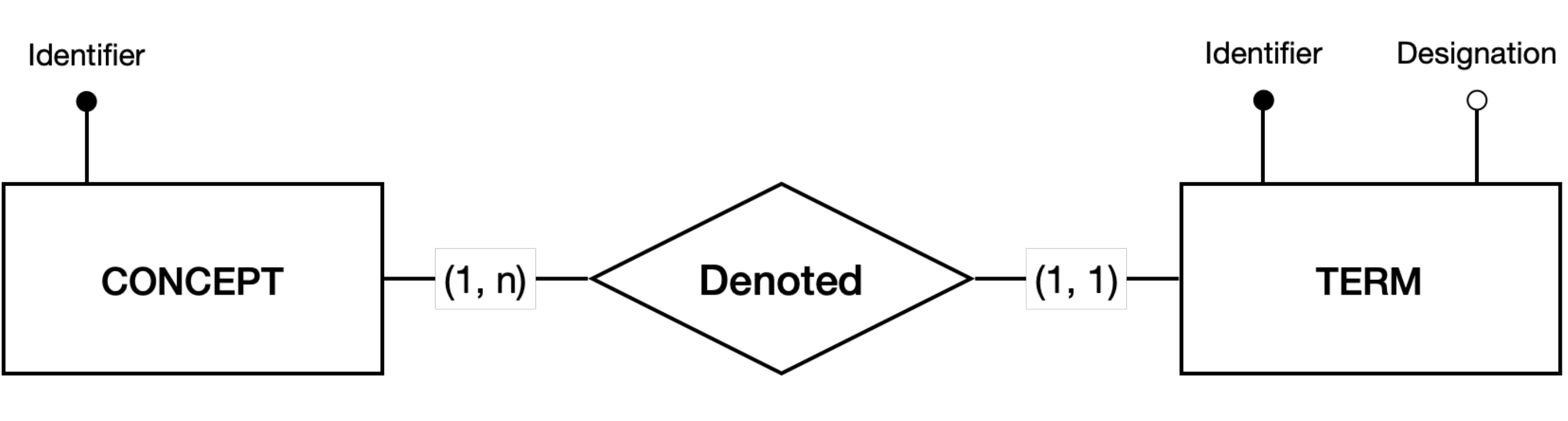}

\caption{ER diagram for concepts denoted by terms}
\label{fig:er_concept}
\end{center}
\end{figure}

The previous paragraph describes a particular requirement of the domain of interest where a concept can be designated by more than one term. The corresponding ER diagram is shown in Figure~\ref{fig:er_concept}. There are two entities: the entity \textsc{concept} which represents the class of all the elements (the concepts) that are uniquely specified by an identifier, and the entity \textsc{term} which represents the class of elements (the terms) that have a unique identifier and a designation, i.e. the sequence of characters that compose the (multi-)word that designates the term.
The association Denoted describes the relationship between pairs of elements (concept, term). We have an additional information in this diagram, a pair of numbers that indicates a constraint on the participation of an entity to an association. This information is called cardinality and it tells the minimum and the maximum number of times that an element of an entity can participate to the relationship.
In this case, on the segment that connects \textsc{concept} to Denoted we have a cardinality $(1, n)$ which means that an element of \textsc{concept} must be denoted by at least $1$ term and at most by $n$ terms (we use $n$ when there are no specified numbers). On the other side of the association, each element of \textsc{term} must be associate to 1 and only 1 element of \textsc{concept}.

This simple example already contains some interesting food for thoughts: while the cardinality of the \textsc{concept} derives directly from the textual description of the problem, the cardinality $(1, 1)$ of \textsc{term} was our implicit choice to say that \textit{homographs} must be uniquely identified even though the designation is the same. For example, the sequence of characters \textit{bank} is the value of the attribute designation of two different (and uniquely identified) elements of \textsc{term} that refer to the concept that describes the side of a river and the other concept that refers to the financial institution.

\begin{figure}[t]
\begin{center}

\includegraphics[width=\textwidth]{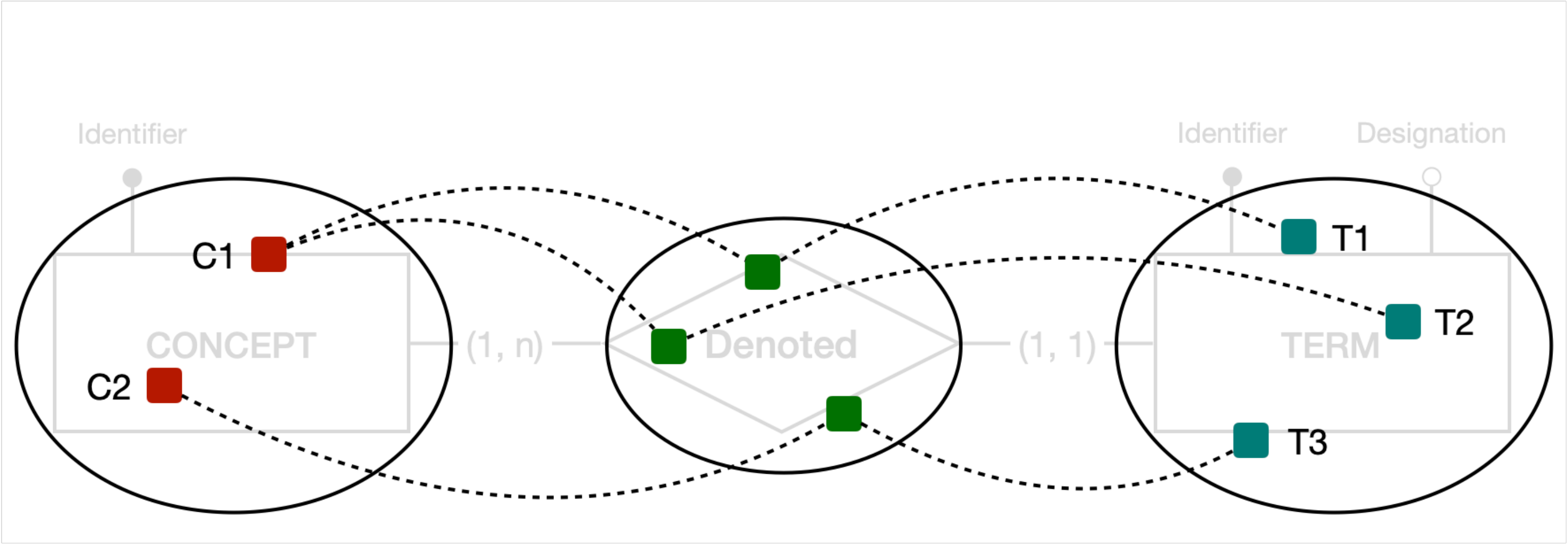}

\caption{Extensional representation of concepts denoted by terms}
\label{fig:er_concept_extensional}
\end{center}
\end{figure}

To conclude this section, we show in Figure~\ref{fig:er_concept_extensional} the extensional representation (sets of elements are explicitly depicted) of the ER diagram. In the background and greyed out, we have the same ER diagram of Figure~\ref{fig:er_concept}, in the foreground, we have the sets (classes) of elements and the explicit links to the association. Concept \textsc{c1} is denoted by two terms, \textsc{t1} and \textsc{t2}, while Concept \textsc{c2} is denoted by term \textsc{t3} only.

\section{From The General Theory of Terminology to Terminological Data Modeling Approach}\label{sec:model}

Despite the title of this section, it is not our intent to enter into the debates and the criticisms to the General Theory of Terminology of Eugene W\"uster. 
We will rather propose an analysis of the requirements of the different theories and approaches to terminology summarised in~\cite{cabre_2003}, that starts from the theory proposed by W\"uster and the design of its corresponding ER diagram.

\subsection{Onomasiological approach}

The main aspect of the General Theory of Terminology proposed by W\"uster  revolves around the notion of concepts, while terms are a sort of ``consequence'' of the organization of the concepts. In the onomasiological approach, concepts are delineated by delimiting characteristics~\cite{suonuuti_2012} while terms are assigned permanently to concepts; in particular, term denotes only one concept. Concepts can be placed in a concept system that describes how concepts are related to each other. 

If we analyze these requirements and compare them with the previous example shown in Figure~\ref{fig:er_concept}, there are at least two new classes of objects involved in the problem: the characteristics of a concept, and the relations among concepts to describe the conceptual system. If we carry out a short analysis of requirements, we find that, according to the ISO 1087:2019~\cite{iso_1087_2019}, there are `type of characteristics' that are categories of characteristics that are grouped (for the purpose of the terminological analysis), essential, non-essential, and delimiting characteristics. Moreover, there are `hierarchical' or `generic' relations among concepts.

\begin{figure}[t]
\begin{center}

\includegraphics[width=\textwidth]{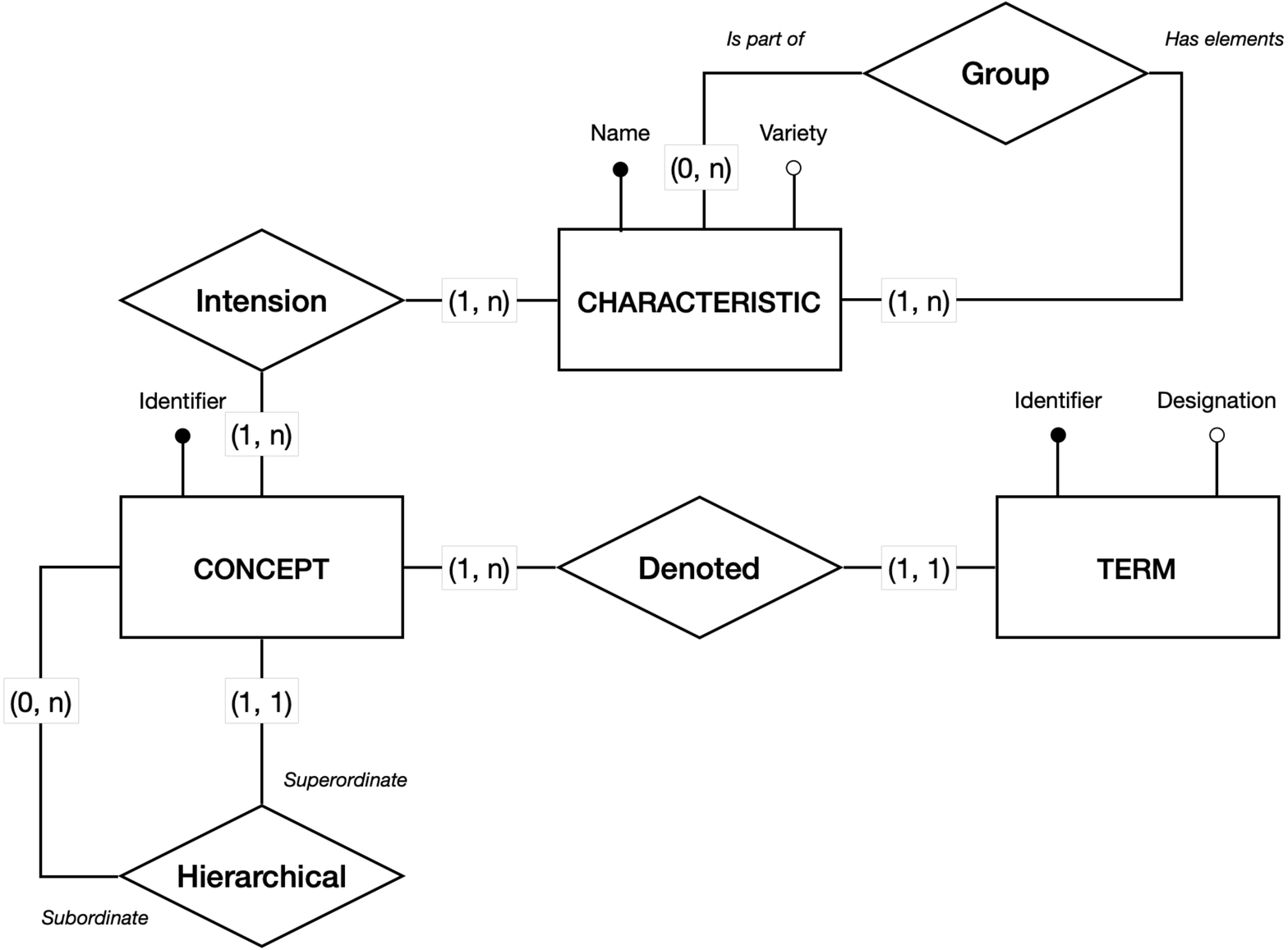}

\caption{Simplified ER diagram for the onomasiological, semasiological, ontoterminological, frame-based approaches.}
\label{fig:er_onomsaiological}
\end{center}
\end{figure}

In Figure~\ref{fig:er_onomsaiological}, we show the resulting ER diagram that includes both the characteristics of a concept and the hierarchical relations among them. For space reasons, the generic relations are not displayed (it would be a similar association that connects \textsc{concept} with itself). The entity \textsc{characteristic} represents the class of elements identified by the `name' (we assume that the name of the characteristic is unique) and has an attribute `variety' to describe whether the element belongs to the type, essential, non-essential, or delimiting characteristic. The association Group describes the fact that: 1) on the one hand, an element of \textsc{characteristic} can be part (see here that it is not mandatory for a characteristic to be part of a group) of one or more types of characteristics (the segment \textit{is part of}), 2) on the other hand, the segment \textit{Has elements} indicate an element `type of characteristic' that can group one or more characteristics. The association Hierarchical models the hierarchy of sub-ordinate and super-ordinate concepts. The cardinality show that an element of \textsc{concept} can (it is optional) have sub-ordinate concepts, while every concept must have at most one superordinate concept (it is not the goal of this paper to discuss whether a concept can belong to different hierarchies or the situation where the concept-of-concepts has as a super-ordinate concept itself).

\subsection{Semasiological approach}

As Santos and Costa argue in~\cite{santos_costa_2015}:
\begin{quote}
That the concept is a central element is not, for us, the question. The question is, knowing if in terminological work it is always the starting point as W\"uster argues or if, depending on the circumstances, it can be the point of arrival.
\end{quote}
The semasiological approach stems from a linguistic analysis of the texts to study the language `in action' and identify terms before concepts. Despite being clearly a sort of `opposite direction' flow of analysis (it is not our intent to oppose the two approaches), the relative conceptual data model is pretty much identical to the one shown in Figure~\ref{fig:er_onomsaiological}. We may want to add additional entities and relationships to document, for example, the \textsc{text} and the \textsc{collection} from which the term was extracted, as well as new types of relationships such as `connected to', `consists of', 'is a', to build the concept map of a domain of knowledge (we would strongly suggest to do it in a real case scenario,). Nevertheless, we witness the fact that the core structure of the conceptual data model is the same to that of the onomasiological one.

\subsection{Ontoterminological approach}

Continuing with the idea of this dichotomy of flow of analysis about how to build knowledge, we analyze the paradigm called ontoterminology~\cite{roche_2012}: ``a terminology whose conceptual system is a formal ontology relying on epistemological principles''. The aim of this paradigm derives from the need to make the operationalization on terminology easier and formally correct, especially from a description logic point of view.

One of the most important features of this paradigm is the double semantic triangle that maps the linguistic knowledge extracted from text to the ontologies derived from domain experts. As a consequence, the definition for a term and for a concept are different in the sense that the former derives from the natural language explanation, while the latter is a formal definition in terms of the characteristics. 

Once again, whereas this approach is different from the other two,
the conceptual data model is still the same. There is one important aspect that we must take into consideration, though: the definition of a term should be included (probably as an attribute of \textsc{term}) to the ER diagram in Figure~\ref{fig:er_onomsaiological}.

\subsection{Frame-based approach}

A similar perspective on the problem of knowledge driven versus corpus driven terminology and on how to enhance the linguistic properties of an ontology is provided by frame semantics~\cite{fillmore_1985}. In the frame based terminology approach~\cite{faber_2015}, the notion of semantic frames which is a structural background knowledge able to construct the meaning of terms is applied. A semantic frame ``models a given situation; situations comprise participants, props, and other conceptual elements, which constitute its frame elements''\cite{lhomme_2018}.

Even in this approach, the core structure of the conceptual data model remains the same. Frames and frame elements could be represented as additional entities related to terms, and their relationships, that describe framework for the definition and representation of specialized knowledge units.

\section{Final Considerations and Future Work}\label{sec:final}

In the previous section, we have presented an investigation of how to model the different theories in terminology. We do not intend to create a new integrated theory, we rather want to study whether a unified conceptual data model that accepts all the different facets that experts in terminology have proposed exists. In this regard, our proposal is  similar to the Cabr\'e's``theory of doors''\cite{cabre_2003} where the terminological unit is the object under observation that can be accessed from different perspectives (the ``doors'') in such a way that the central object, the terminological unit, is directly addressed whether starting from the concept or the term (or even `the situations' as described in her work). In our design of the conceptual data model, the core data structure is the object that can be accessed and the ``approaches'' are only a different view of the same data. Of course, this does not mean that different approaches will produce different data for the same object under study, but the idea is that different views can enhance the information about the term (ontological, linguistic, practical information) or simply access portions of data.

Moreover, the ER model can be used as a basis for unification of different views of data: the network model, the relational model, and the entity set model. This fact will be very useful for what comes next. The positive consequence of this homogeneous conceptual data model is that we can decide what `direction' we want (or need) for the implementation of the structure that will represent the data and how we want to export them.
For example, we can:
\begin{itemize}
\item transform the ER diagram into the corresponding relational database and export the standard TermBase Exchange (TBX) file~\cite{vezzani_dinunzio_2020};
\item provide an experimental infrastructure for the study of the properties of terms and their relations with distributional semantics\cite{bonato_et_al2021}, or the study of the quality of machine translations~\cite{cambedda_et_al_2021};
\item make a comparative qualitative quantitative analysis of the use of specialized languages in narrative~\cite{vezzani_dinunzio_2019};
\item study the performance of query variation for technology assisted medical systematic review systems~\cite{dinunzio_vezzani_2021}.
\end{itemize}

In addition, we may want to increase the level of interoperability by exposing the data with a Linked Open Data approach. Since the ER diagram can be mapped to a graph model, we can directly map the overall conceptual data model to a graph and expose it, for example in RDF \cite{dibuccio_et_al_2013,dibuccio_et_al_2014}.

\bibliographystyle{splncs04}
\bibliography{toth_2021}

\begin{thebibliography}{10}
\providecommand{\url}[1]{\texttt{#1}}
\providecommand{\urlprefix}{URL }
\providecommand{\doi}[1]{https://doi.org/#1}

\bibitem{bonato_et_al2021}
Bonato, V., Di~Nunzio, G.M., Vezzani, F.: Preliminary considerations on a
  systematic approach to semic analysis: The case study of medical terminology.
  Umanistica Digitale (10),  211--234 (Jan 2021).
  \doi{10.6092/issn.2532-8816/12621},
  \url{https://umanisticadigitale.unibo.it/article/view/12621}

\bibitem{dibuccio_et_al_2013}
Buccio, E.D., Nunzio, G.M.D., Silvello, G.: A curated and evolving linguistic
  linked dataset. Semantic Web  \textbf{4}(3),  265--270 (2013).
  \doi{10.3233/SW-2012-0083}, \url{https://doi.org/10.3233/SW-2012-0083}

\bibitem{dibuccio_et_al_2014}
Buccio, E.D., Nunzio, G.M.D., Silvello, G.: A linked open data approach for
  geolinguistics applications. Int. J. Metadata Semant. Ontologies
  \textbf{9}(1),  29--41 (2014). \doi{10.1504/IJMSO.2014.059125},
  \url{https://doi.org/10.1504/IJMSO.2014.059125}

\bibitem{dibuono_et_al_2020a}
di~Buono, M.P., Cimiano, P., Elahi, M.F., Grimm, F.: Terme-{\`{a}}-llod:
  Simplifying the conversion and hosting of terminological resources as linked
  data. In: Proceedings of the 7th Workshop on Linked Data in Linguistics,
  LDL@LREC 2020, Marseille, France, May 2020. pp. 28--35 (2020),
  \url{https://aclanthology.org/2020.ldl-1.5/}

\bibitem{cabre_2003}
Cabr{\'e}~Castellv{\'\i}, M.T.: Theories of terminology: Their description,
  prescription and explanation. Terminology. International Journal of
  Theoretical and Applied Issues in Specialized Communication  \textbf{9}(2),
  163--199 (2003). \doi{https://doi.org/10.1075/term.9.2.03cab},
  \url{https://www.jbe-platform.com/content/journals/10.1075/term.9.2.03cab}

\bibitem{cambedda_et_al_2021}
Cambedda, G., Di~Nunzio, G.M., Nosilia, V.: A study on automatic machine
  translation tools: A comparative error analysis between deepl and yandex for
  russian-italian medical translation. Umanistica Digitale (10),  139--163 (Jan
  2021). \doi{10.6092/issn.2532-8816/12631},
  \url{https://umanisticadigitale.unibo.it/article/view/12631}

\bibitem{chen_1976}
Chen, P.P.S.: The entity-relationship model---toward a unified view of data.
  ACM Trans. Database Syst.  \textbf{1}(1),  9--36 (Mar 1976).
  \doi{10.1145/320434.320440}, \url{https://doi.org/10.1145/320434.320440}

\bibitem{faber_2015}
Faber, P.: Frames as a framerowk for terminology. In: Handbook of terminology,
  vol.~1, p.~153. John Benjamins Publishing Company, Amsterdam (2015)

\bibitem{fillmore_1985}
Fillmore, C.J.: Frames and the semantics of understanding. Quaderni di
  Semantica  \textbf{6}(2),  222--254 (1985)

\bibitem{iso_1087_2019}
{International Organization for Standardization}: Terminology work and
  terminology science --- vocabulary.
  https://www.iso.org/obp/ui/\#iso:std:iso:1087:en (2019)

\bibitem{lhomme_2018}
L'Homme, M.C.: Maintaining the balance between knowledge and the lexicon in
  terminology: a methodology based on frame semantics. Lexicography
  \textbf{4}(1),  3--21 (2018). \doi{10.1007/s40607-018-0034-1},
  \url{https://doi.org/10.1007/s40607-018-0034-1}

\bibitem{dinunzio_vezzani_2021}
Nunzio, G.M.D., Vezzani, F.: {IMS-UNIPD} @ {CLEF} ehealth task 2: Reciprocal
  ranking fusion in {CHS}. In: Proceedings of the Working Notes of {CLEF} 2021
  - Conference and Labs of the Evaluation Forum, Bucharest, Romania, September
  21st - to - 24th, 2021. pp. 775--779 (2021),
  \url{http://ceur-ws.org/Vol-2936/paper-64.pdf}

\bibitem{reineke_2014}
Reineke, D.: {TBX between termbases and ontologies}. In: {Terminology and
  Knowledge Engineering 2014}. Berlin, Germany (Jun 2014),
  \url{https://hal.archives-ouvertes.fr/hal-01005838}

\bibitem{reineke_romary_2019}
Reineke, D., Romary, L.: {Bridging the gap between SKOS and TBX}. {edition -
  Die Fachzeitschrift f{\"u}r Terminologie}  \textbf{19}(2) (Nov 2019),
  \url{https://hal.inria.fr/hal-02398820}

\bibitem{roche_2012}
Roche, C.: Ontoterminology: How to unify terminology and ontology into a single
  paradigm. In: Proceedings of the Eighth International Conference on Language
  Resources and Evaluation, {LREC} 2012, Istanbul, Turkey, May 23-25, 2012. pp.
  2626--2630 (2012),
  \url{http://www.lrec-conf.org/proceedings/lrec2012/summaries/567.html}

\bibitem{santos_costa_2015}
Santos, C., Costa, R.: Domain specificity. In: Handbook of terminology, vol.~1,
  p.~153. John Benjamins Publishing Company, Amsterdam (2015)

\bibitem{dibuono_et_al_2020b}
Speranza, G., di~Buono, M.P., Monti, J., Sangati, F.: From linguistic resources
  to ontology-aware terminologies: Minding the representation gap. In:
  Proceedings of The 12th Language Resources and Evaluation Conference, {LREC}
  2020, Marseille, France, May 11-16, 2020. pp. 2503--2510 (2020),
  \url{https://aclanthology.org/2020.lrec-1.305/}

\bibitem{suonuuti_2012}
Suonuuti, H.: Guide to terminology. Tekniikan sanastokeskus, Helsinki, 2nd edn.
  (2012)

\bibitem{vezzani_dinunzio_2019}
Vezzani, F., Di~Nunzio, G.M.: (not so) elementary, my dear watson! a different
  perspective on medical terminology. Umanistica Digitale  \textbf{3}(6) (Jan
  2019). \doi{10.6092/issn.2532-8816/8632},
  \url{https://umanisticadigitale.unibo.it/article/view/8632}

\bibitem{vezzani_dinunzio_2020}
Vezzani, F., Di~Nunzio, G.M.: Methodology for the standardization of
  terminological resources: Design of trimed database to support multi-register
  medical communication. Terminology. International Journal of Theoretical and
  Applied Issues in Specialized Communication  \textbf{26}(2),  265--297
  (2020). \doi{https://doi.org/10.1075/term.00053.vez},
  \url{https://www.jbe-platform.com/content/journals/10.1075/term.00053.vez}

\end{thebibliography}

\end{document}